\documentclass[aps,prd,amsmath,amssymb,preprintnumbers,onecolumn,11pt,nofootinbib]{revtex4}

\usepackage[a4paper,left=20mm,right=20mm,top=25mm,bottom=25mm]{geometry}
\usepackage{amsmath,amssymb}
\usepackage{mathrsfs}
\usepackage{graphicx}
\usepackage{xcolor}
\usepackage{subfigure}
\usepackage{fancyhdr}
\usepackage{multirow}
\usepackage{float}
\usepackage{epsfig}
\usepackage{amsfonts}
\usepackage{bm}
\usepackage{hyperref}
\usepackage{xcolor}

\definecolor{CP3}{cmyk}{0,0.88,0.77,0.40}

\newcommand{\cL}{\mathcal{L}}

\newcommand{\al}{\alpha}
\newcommand{\bt}{\beta}

\newcommand{\pa}{\partial}

\newcommand{\be}{\begin{equation}}
\newcommand{\ee}{\end{equation}}
\newcommand{\ba}{\begin{eqnarray}}
\newcommand{\ea}{\end{eqnarray}}

\renewcommand{\(}{\left(}
\renewcommand{\)}{\right)}
\renewcommand{\[}{\left[}
\renewcommand{\]}{\right]}

\newcommand\nb{\bar{N}}

\newcommand\difl{\tilde{\lambda}}

\begin{document}

\title{Observational predictions of inflationary model in spatially covariant gravity with two tensorial degrees of freedom for gravity}

\author{Saikat Chakraborty}
\email{saikat.ch@nu.ac.th}
\affiliation{The Institute for Fundamental Study “The Tah Poe Academia Institute”, Naresuan University, Phitsanulok 65000, Thailand}
\affiliation{Center for Space Research, North-West University, Mahikeng 2745, South Africa}

\author{Khamphee Karwan}
\email{khampheek@nu.ac.th}
\affiliation{The Institute for Fundamental Study “The Tah Poe Academia Institute”, Naresuan University, Phitsanulok 65000, Thailand}

\author{Jakkrit Sangtawee}%
\email{jakkrits60@nu.ac.th}
\affiliation{The Institute for Fundamental Study “The Tah Poe Academia Institute”, Naresuan University, Phitsanulok 65000, Thailand}

\date{\today}

\begin{abstract}

We study the inflationary model constructed from a Spatially Covariant Gravity (SCG). The Lagrangian for the SCG in our consideration is expressed as the polynomial of irreducible SCG monomials where the total number of derivatives of each monomial is two,
and the theory propagates two tensorial degrees of freedom of gravity up to the first order in cosmological perturbations.
The condition for having two tensorial degrees of freedom studied earlier in literature for such theories is derived in vacuum. 
We extend the condition for having two tensorial degrees  of freedom to the case where a scalar field is included by imposing a gauge-fixing. 
We apply  the resulting  SCG to describe inflationary universe.
The observational predictions such as the scalar spectral index and tensor-to-scalar ratio from this model are investigated.
We find that the tensor-to-scalar ratio in this model can either be in the order of unity or be small depending on the parameter of the model.

\end{abstract}


\maketitle

\section{Introduction}

The cosmic inflation is a  framework for describing the early universe \cite{Starobinsky:1980, Guth:1981, Linde:1981, Albrecht:1982}.
It can address shortcomings of Big Bang cosmology,
and can also provide a mechanism for generating primordial fluctuations which are seeds of the large-scale structures in the present universe.
To drive the inflationary dynamics, an extra degree of freedom has to be added to the underlying theory.
This extra degree of freedom could be a minimally coupled scalar field namely the inflaton or could be a scalar degree of freedom for gravity appearing in modified theories of gravity such as scalar-tensor theories of gravity or $f(R)$ gravity (see e.g. \cite{Brooker:2016oqa}).

In general, modified theories of gravity can be constructed by adding extra degrees of freedom to gravity.
However, it is also possible to construct modified theories of gravity that have two tensorial degrees of freedom for gravity as in General Relativity (GR).
These modified theories of gravity can be constructed by breaking temporal diffeomorphism invariance in the context of the 3+1 formalism, i.e., the Hamiltonian constraint of such theories is not a time-translation generator \cite{Carballo-Rubio:2018czn, Gao:2014soa, Gao:2014fra}.
The class of such modified theories of gravity is called Minimally Modified Gravity (MMG) theories \cite{Lin:2017oow, Aoki:2018zcv, Aoki:2018brq, Mukohyama:2019unx, DeFelice:2020eju, Gao:2019twq, Yao:2020tur, Hu:2021yaq}.
The MMG theories can be constructed by imposing a condition on the form of Lagrangian or Hamiltonian of theories such that the theories have two tensorial propagating degrees of freedom for gravity \cite{Lin:2017oow, Mukohyama:2019unx}.
It has been shown that the condition for a gravity theory to be an MMG in a vacuum is not sufficient when matter appears in the theory. This is because the Hamiltonian constraint is deformed and the deformed Hamiltonian constraint is not first class anymore \cite{Carballo-Rubio:2018czn, Aoki:2018zcv}.
Hence, the application of the MMG theories to cosmology, where we usually always have matter, requires the additional gauge-fixing condition to ensure that only two tensorial gravitational degrees of freedom propagate in the theories \cite{Aoki:2018zcv, Aoki:2020oqc, Carballo-Rubio:2018czn}.
Cosmological consequences of the MMG theories have been studied in \cite{Aoki:2020oqc, Sangtawee:2021mhz, Ganz:2022iiv, Ganz:2022zgs}.

Possible construction of MMG theories can be started from SCG \cite{Gao:2019twq, Gao:2019lpz, Hu:2021yaq}. 
Due to the explicit breaking of the temporal diffeomorphism, an SCG, in general, propagate an additional scalar degree of freedom \footnote{This can be made apparent by using an inverse Stuckelberg trick.}
The condition for having two tensorial degrees of freedom has been investigated from the second order action for cosmological perturbations in vacuum \cite{Hu:2021yaq}.
In this work, we extend the analysis in \cite{Hu:2021yaq} by adding a scalar field in the consideration, which is a more cosmologically relevant scenario.
We then use this model to describe an inflationary universe and compute observational predictions.

This work is organized as follows.
In Sec.~\ref{sec:2}, we present the action for spatially covariant gravity in our analysis.
We discuss a possible gauge-fixing in Lagrangian formalism in Sec.~\ref{sec:3}.
The actions for the first and second orders perturbations are computed in Secs.~\ref{sec:4} and \ref{sec:5}.
In Sec.~\ref{sec:6}, we apply the results from Secs.~\ref{sec:4} and \ref{sec:5} to inflationary universe.
We conclude in Sec.~\ref{sec:7}.

\section{Spatially covariant gravity } 
\label{sec:2}
We consider the Lagrangian for gravity in the form of a polynomial of SCG monomials where the total number of derivatives of each monomial is two \cite{Hu:2021yaq, Gao:2020yzr}:
\be
\cL = c_{1} K_{ij}K^{ij}+ c_{2} K^{2} + c_{3} R\,,
\label{lang}
\ee
where the coefficients $c_1, ..., c_{3}$ depend on time $t$ and the lapse function $N$,
and $R$ is the 3-dimensional Ricci scalar.

The extrinsic curvature can be computed as
\be
K_{ij}=\frac{1}{2N}\left(\frac{\partial h_{ij}}{\partial t} - D_i N_j - D_j N_i\right)\,,
\ee
where $N^i$ is the shift vector, $D_i$ is the covariant derivative compatible with the metric $h_{ij}$.
The condition for which this theory propagates only two tensorial degrees of freedom in a vacuum has been investigated using the perturbative approach in \cite{Hu:2021yaq}. 
The idea is that an additional propagating scalar degree of freedom, if absent from the theory, should not show up in any order of perturbations. Using this approach, although one may not be able to find the exact condition for the absence of the additional propagating scalar degree of freedom, one may find at each order the condition such that the degree of freedom is not propagating at that order. In particular, it was shown that, up to the linear order in perturbations, the theory given by the Lagrangian \eqref{lang} propagates only two tensorial degrees of freedom, provided the following condition is satisfied
\be\label{condition_vacuum}
b_2'' = - \frac{2b_2'(b_2 - b_2')}{b_2}\,,
\ee
where $b_2=c_1+3c_2$

In this work, we are interested in the inflationary model constructed from the above theory up to first order in perturbations.
It is well known that the condition for having two tensorial degrees of freedom for gravity in MMG theory that is constructed in a vacuum is no longer sufficient to eliminate the additional degrees of freedom for gravity arising when a matter is added to the theory. 
The reason is that the extended Hamiltonian constraint of the theory when a matter is added is no longer first class anymore \cite{Aoki:2020oqc}. A possible approach for eliminating such an extra degree of freedom is to impose additional gauge-fixing conditions at the vacuum level, thus splitting the first class constraint into two second class constraints, and then couple matter minimally.
One, therefore, needs to work with this specific choice of gauge or equivalently hypersurface foliations \cite{Aoki:2020oqc}.

To compute the perturbed actions of gravity and scalar field,
the lapse function, shift vector and the 3-dimensional metric are decomposed into the background and perturbed parts as \cite{Aoki:2020oqc}
\begin{eqnarray}
N & = & \bar{N} (1 + \al)\,,
\label{lapse_xpn}\\
N_{i} & = & \bar{N}\,\pa_{i}\bt\,,
\label{shift_xpn}\\
h_{ij} & = & a^{2}e^{2\zeta}\delta_{ij}\,,
\label{sptmetric_xpn}
\end{eqnarray}
where $a$ is a scale factor. 
Note that, in general, the scalar perturbed spatial metric $h_{ij}$ should be characterized by two scalar perturbation quantities, instead of the single quantity $\zeta$. However, we have exploited the spatial diffeomorphism of the theory to fix the spatial gauge and eliminate the other perturbation quantity.

We consider a scalar field with Lagrangian
\be
\cL_\phi = X - V(\phi)\,,
\ee
$X \equiv - \frac 12 \partial_\mu\phi \partial^\mu \phi$ being the kinetic part. The total action is
\be
S = \int dt d^3 x  N \sqrt{h} \(\cL + \cL_\phi\)\,,
\label{act-cf}
\ee
where $h$ is the determinant of $h_{ij}$ and $\cL$ is given by Eq.~(\ref{lang}).
The kinetic term of a scalar field can be decomposed as
\be
X = \frac 1{2 N^2} \[\(\frac{d}{d t}(\phi + \pi)\)^2 - 2 N^i \partial_i \pi \frac{d \phi}{d t} - \(N^2 h^{ij} - N^iN^j\) \partial_i\pi \partial_j\pi\]\,,
\label{kin}
\ee
where $\phi$ and $\pi$ denote the homogeneous background part of the scalar field and the scalar field perturbation respectively.

\section{Fixing a gauge}
\label{sec:3}

To investigate inflation in an MMG we need to include an inflaton scalar field in the picture. However, an MMG theory that is constructed in a vacuum will no longer be an MMG once a scalar field is added. Even in the gravity sector, an additional propagating scalar degree of freedom will show up \footnote{This is on top of the scalar degree of freedom brought by the inflation scalar field itself.}.
This additional degree of freedom for gravity can be eliminated by fixing a gauge. We take this motivation from the Hamiltonian construction of MMG theories, namely the so-called $f(\mathcal{H})$ theories \cite{Mukohyama:2019unx}.

In Hamiltonian formulation,
a gauge fixing condition can be imposed  by adding the Lagrange multiplier term in the form \cite{Aoki:2018zcv,Aoki:2020oqc}
\begin{equation}
H_{\text{gf}}=\int d^{3}x\sqrt{h}\,\lambda^{i}\partial_{i}\left(\frac{\Pi}{\sqrt{h}}\right)\,,
\label{hg}
\end{equation}
 where $\lambda^{i}$ is a Lagrange multiplier,
 and $\Pi$ is the trace of the canonical momentum conjugate to $h_{ij}$ defined by
 \be
 \Pi^{ij} \equiv \frac{\delta}{\delta \dot{h}_{ij}}\int d^3 x \cL\,.
 \ee
In the above equation,
$\cL$ is defined in Eq.~(\ref{lang}).
Eq.~(\ref{hg}) fixes a gauge by choosing a hypersurfaces of uniform $\Pi / \sqrt{h}$, i.e., $\partial_i (\Pi / \sqrt{h}) = 0$.
From \cite{Aoki:2020oqc},
$\Pi_{ij}$ is given by
\begin{equation}
\Pi_{ij}
=\frac{{\sqrt{h}}}{f'(\mathcal{H}_{0})}\(K_{ij}-Kh_{ij}
- \frac 1N h_{ij}D_{k}\lambda^{k}\)\,,
\end{equation}
where $f(\mathcal{H}_{0})$ is an arbitrary function of a Hamiltonian constraint in GR $\mathcal{H}_{0}$,
while a prime over $f$ denotes derivative with respect to  $\mathcal{H}_{0}$.
Hence, a uniform $\Pi/ \sqrt{h}$ hypersurface corresponds to
\be
0 = \partial_i \(\frac{\Pi}{\sqrt{h}}\)= \partial_i \[\frac{1}{f'(\mathcal{H}_{0})}\(- 2K
- \frac 3N h_{ij}D_{k}\lambda^{k}\)\]\,.
\ee
The above gauge fixing condition inspires us to add to our Lagrangian \eqref{lang} a Lagrange multiplier term of the form 
\be
L_{\text{gf}}=\int d^{3}x\sqrt{h}\,\lambda^{i}\partial_{i}K\,,
\label{lg}
\ee
where $\lambda^{i}$ is a Lagrange multiplier.
Adding the above Lagrange multiplier term in Eq.~(\ref{act-cf}),
we get
\be
S = \int dt d^3 x  \sqrt{h} \(N \cL + N \cL_\phi + \lambda^{i}\partial_{i}K\)\,.
\label{act-cfg}
\ee
Varying this action with respect to $\lambda^{i}$,
we obtain a constraint equation
\be
\partial_{i}K = 0\,.
\label{k-uniform}
\ee
This is the gauge condition we work with. In other words, we choose to work in the hypersurface of uniform $K$.
The momentum conjugated to $h_{ij}$ is
\be
\Pi_{ij} = \sqrt{h}\(
c_{1} K_{ij} + c_{2} K h_{ij} - \frac 1{2 N} D_i \lambda^i h_{ij} 
\)\,,
\label{pij-xi}
\ee
and therefore
\be
\frac{\Pi}{\sqrt{h}} = (c_{1} + 3 c_{2} ) K - \frac 3{2 N} D_i \lambda^i\,.
\ee
We see that the condition $\partial_i K = 0$ can correspond to $\partial_i (\Pi/\sqrt{h}) =0$ for a specific condition on the coefficients $c_{1}$ and $c_{2}$.
However, we do not impose such condition on the coefficients in our consideration.
It will be seen in the following analysis that
the considered SCG theory will propagate two tensorial degrees of freedom for gravity up to the first order perturbations if the gauge-fixing in Eq.~(\ref{lg}) is imposed.

\section{Action for the first order perturbations }
\label{sec:4}

Expanding the action in Eq.~(\ref{act-cfg}) up to the first order in perturbation,
we get
\be
S_1 = \int dt d^3 x\bar{N} a^3 \cL_1\,,
\ee
where
\ba
\cL_1&=&
- \frac{1}{2} \bigl(6 b_2 H^2 - 6 b_2' H^2 + \dot{\phi}^2 + 2 V\bigr)\alpha
 - \frac{3}{2} \bigl(6 b_2 H^2 -  \dot{\phi}^2 + 2 V + 4 H \dot{b_2} + 4 b_2 \dot{H}\bigr)\zeta	\nonumber\\
 &&- (V_\phi + 3 H \dot{\phi} + \ddot{\phi})\pi + \mbox{divergence terms }\,.
\label{l1r}
\ea
Note that the  Lagrange multiplier term corresponds to the divergence term at the first order in perturbation.
We follow \cite{Hu:2021yaq} to define
\ba
\dot{X} &\equiv & \frac{1}{\bar{N}}\frac{\partial X}{\partial t}\,,
\qquad H \equiv \frac{\dot{a}}{a}\,,
\nonumber\\
f' &\equiv & \bar{N}\left.\frac{\partial f}{\partial N}\right|_{N=\bar{N}}\,,
\qquad 
f''\equiv\bar{N}^{2}\left.\frac{\partial^{2}f}{\partial N^{2}}\right|_{N=\bar{N}}\,.
\label{notations}
\ea
Since the action for the first order perturbation vanishes,
the coefficients of $\al, \zeta$ and $\pi$ vanish yielding the following evolution equations for the background
\ba
0&=& 6 b_2 H^2 - 6 b_2' H^2 + \dot{\phi}^2 + 2 V\,,
\label{eom-b1}\\
0 &=& 6 b_2 H^2 -  \dot{\phi}^2 + 2 V + 4 H \dot{b_2} + 4 b_2 \dot{H}\,.
\label{eom-b2}\\
0 &=& V_\phi + 3 H \dot{\phi} + \ddot{\phi}\,,
\label{eom-b3}
\ea
where $V_\phi \equiv d V / d \phi$.

\section{Action for the second order perturbations and condition for having two degree of freedom for gravity}
\label{sec:5}

The dynamics of the first order perturbations are governed by the evolution equations obtained from the action for the second order perturbations.
To compute the action for the second order perturbations,
we integrate by part the Lagrange multiplier term such as
\be
S = \int dt d^3 x  \sqrt{h} \(N \cL + N \cL_\phi - D_i \lambda^{i} K\)\,.
\label{act-cfg-bp}
\ee
Expanding the above action to the second order in perturbation,
we get
\be
S_2 = \int dt d^3 x \nb a^3 \cL_2\,,
\label{act2}
\ee
where
\ba
\cL_2 &=&
\frac{1}{2} \Bigg(3 \alpha^2 b_2'' H^2 - 2 \alpha^2 V - 2 \alpha \pi V_\phi -  \pi^2 V_{\phi \phi} - 6 \pi V_\phi \zeta + 27 b_2 H^2 \zeta^2 - 9 V \zeta^2 + 9 X \zeta^2 + \dot{\pi}^2 - 12 \alpha \Omega_1 H \dot{\zeta} \nonumber\\
&& + 36 b_2 H \zeta \dot{\zeta} + 6 b_2 \dot{\zeta}^2 + \frac{4 \alpha \Omega_1 H \beta {}^{,i}{}_{,i}}{a^2} -  \frac{4 b_2 \dot{\zeta} \beta {}^{,i}{}_{,i}}{a^2} + \dot{\phi} \(-2 (\alpha - 3 \zeta) \dot{\pi} + \frac{2 \pi \beta {}^{,i}{}_{,i}}{a^2}\) -  \frac{8 \alpha c_{3} \zeta {}^{,i}{}_{,i}}{a^2} \nonumber\\
&& - \frac{8 \alpha c_{3}' \zeta {}^{,i}{}_{,i}}{a^2} -  \frac{\pi {}_{,i} \pi {}^{,i}}{a^2} + \frac{4 c_{3} \zeta {}_{,i} \zeta {}^{,i}}{a^2} + \frac{2 \Omega_2 \beta {}^{,i}{}_{,i} \beta {}^{,k}{}_{,k}}{a^4}\Bigg)
\nonumber\\
&& + \difl \(3 \alpha H - 3 \dot{\zeta} + \frac{\beta {}^{,k}{}_{,k}}{a^2}\)\,,
\ea
where the last line comes from the Lagrange multiplier term, 
the derivative with respect to the comoving spatial coordinates is denoted by  $\partial / \partial x^i \equiv {}_{, i}$,
and $\difl \equiv (\partial \lambda^i / \partial x^i) / \bar{N}$ and $\Omega_1 \equiv b_2 - b_2'$.

In the last line, we have integrated by part the term $\Gamma_{ij}^i \lambda^j$ such that the result depends only on $\difl$.
Note that, due to its vectorial nature and the background being isotropic, $\lambda^i$ vanishes at the background level and we can treat $\difl$ as a first-order perturbed quantity. Varying the action (\ref{act2}) with respect to $\al$, $\beta_{,i}^{, i}$ and $\difl$,
we get
\ba
 0 &=& 3 \alpha b_2'' H^2 - 2 \alpha V -  \pi V_\phi -  \dot{\phi} \dot{\pi} - 6 \Omega_1 H \dot{\zeta} + \frac{2 \Omega_1 H \beta {}^{,i}{}_{,i}}{a^2} -  \frac{4 c_{3} \zeta {}^{,i}{}_{,i}}{a^2} -  \frac{4 c_{3}' \zeta {}^{,i}{}_{,i}}{a^2}
\nonumber\\
&& 
+ 3 \difl H\,,
\label{eq-al}\\
0 &=& \frac{2 a^2 \alpha \Omega_1 H + a^2 \pi \dot{\phi} - 2 a^2 b_2 \dot{\zeta} + 2 \Omega_2 \beta {}^{,i}{}_{,i}}{a^4} + \frac{\difl}{a^2}\,,
\label{eq-bt}\\
0 &=& 3 \alpha H - 3 \partial_t \zeta + \frac{\beta {}^{,i}{}_{,i}}{a^2}\,,
\label{eq-lm}
 \ea
 where $\Omega_2 \equiv b_2 - 2 c_{2}$.
Eqs.~(\ref{eq-al}) -- (\ref{eq-lm}) can be solved to obtain
\ba
\beta_{,i}^{, i} &=& - \frac{3 \bigl(a^2 H \pi V_\phi + a^2 H \dot{\phi} (3 H \pi + \dot{\pi}) + a^2 (-6 b_2 H^2 - 3 b_2'' H^2 + 12 \Omega_1 H^2 + 2 V) \dot{\zeta} + 4 (c_{3} + c_{3}') H \zeta {}^{,i}{}_{,i}\bigr)}{2 (3 b_2'' H^2 - 9 \Omega_1 H^2 + 9 \Omega_2 H^2 - 2 V)}\,,\nonumber\\
\label{p2btSol} \\
\alpha &=& \frac{a^2 \pi V_\phi + a^2 \dot{\phi} (3 H \pi + \dot{\pi}) + 3 a^2 (-2 b_2 + \Omega_1 + 3 \Omega_2) H \dot{\zeta} + 4 c_{3} \zeta {}^{,i}{}_{,i} + 4 c_{3}' \zeta {}^{,i}{}_{,i}}{a^2 (3 b_2'' H^2 - 9 \Omega_1 H^2 + 9 \Omega_2 H^2 - 2 V)}\,,
\label{alSol}\\
\difl &=&
\frac{1}{a^2 \bigl(3 b_2'' H^2 - 2 (6 \Omega_1 H^2 - 9 \Omega_2 H^2 + V)\bigr)}\big[a^2 \dot{\phi} \bigl(\pi (-3 b_2'' H^2 + 6 \Omega_1 H^2 + 2 V) - 2 (\Omega_1 - 3 \Omega_2) H \dot{\pi} \bigr) \nonumber\\
&& + 2 a^2 (b_2 - 3 \Omega_2) \bigl(3 b_2'' H^2 - 2 (3 \Omega_1 H^2 + V)\bigr) \dot{\zeta} - 2 (\Omega_1 - 3 \Omega_2) H \bigl(a^2 \pi V_\phi + 4 (c_{3} + c_{3}') \zeta {}^{,i}{}_{,i}\bigr)\big]\,.
\label{lmSol}
\ea
Substituting Eqs.~(\ref{p2btSol}) -- (\ref{lmSol}) into Eq.~(\ref{act2}),
we can write the kinetic part of the action as
\be
S_2^K = \int dt d^3 x a^3 \nb \dot{\cal V}^{T} {\cal M} \dot{\cal V}\,,
\ee
where $\mathcal{V} \equiv \begin{pmatrix}
  \zeta\\ 
  \pi
\end{pmatrix}$ and the kinetic matrix $\mathcal{M}$ has the elements
\ba
{\cal M}_{11} &=& - \frac{3 (b_2 - 3 \Omega_2) (6 b_2 H^2 + 3 b_2'' H^2 - 6 \Omega_1 H^2 + 2 X)}{3 b_2'' H^2 + 2 (-3 \Omega_1 H^2 + 9 \Omega_2 H^2 + X)}\,,
\label{m11}\\
{\cal M}_{12} &=& {\cal M}_{21} =
\frac{3 (b_2 - 3 \Omega_2) H \dot{\phi}}{3 b_2'' H^2 + 2 (-3 \Omega_1 H^2 + 9 \Omega_2 H^2 + X)}\,,
\label{m12}\\
{\cal M}_{22} &=& 
\frac{3 (b_2'' - 2 \Omega_1 + 6 \Omega_2) H^2}{6 b_2'' H^2 + 4 (-3 \Omega_1 H^2 + 9 \Omega_2 H^2 + X)}\,.
\label{m22}
\ea
When a scalar field is included in an MMG theory,
we expect that the theory should propagate one scalar degree of freedom in addition to two tensorial degrees of freedom.
The scalar degree of freedom  is the degree of freedom coming from the scalar field itself,
while the tensorial degrees of freedom are the degrees of freedom for gravity.
However, in our analysis, both $\zeta$ and $\pi$ can be propagating degrees of freedom because ${\cal M}_{11}$ and ${\cal M}_{22}$ do not vanish in general.
Since only a scalar field is included in the theory,
the existence of two propagating scalar degrees of freedom implies that one of these degrees of freedom could be a degree of freedom for gravity.
Up to the first order in perturbations, the theory could propagate only one scalar degree of freedom
(and therefore the gravitational interaction will be delivered by two tensorial degrees of freedom only), provided one of the eigenvalues of the kinetic matrix vanishes. 
It can be checked that one of the eigenvalues of the kinetic metric vanishes if
\be
b2'' = -2 b_2'\,.
\label{condition}
\ee
Note that this condition is different from the corresponding condition in vacuum \eqref{condition_vacuum}. When a matter is added, the condition for only two propagating gravitational degrees of freedom changes.

Inserting Eq.~(\ref{condition}) in Eqs.~(\ref{m11}) -- (\ref{m22}),
the eigenvectors of the kinetic matrix are
\be
{\cal V}_1 = \(\begin{array}{c}
-\frac{\dot\phi}{H}\\
1
\end{array}\)\,,
\quad
{\cal V}_2 = \(\begin{array}{c}
\frac{H}{\dot\phi}\\
1
\end{array}\)\,.
\ee
Performing a similarity  transformation, we find the
vectors in the $\zeta-\pi$ plane that diagonalizes the kinetic matrix. They can take the form
\be
\(\begin{array}{c}
\tilde{\zeta} \\
\tilde{\pi}
\end{array}\)
= \(\begin{array}{c}
\frac{\dot{\phi}}{H } \pi + \zeta
\\
\pi -  \frac{\dot{\phi}}{H } \zeta
\end{array}\)\,.
\label{simtransf}
\ee
It can be seen that $\tilde{\pi}$ is a gauge invariant variable, but $\tilde\zeta$ is not. The gauge-invariant quantity $\tilde{\pi}$ is defined in the same way as the scalar field perturbation in the uniform 3-curvature gauge.
Nevertheless, we do not work in  a specific gauge because the temporal diffeomorphism is broken.

In what follows, we would work entirely with the variables $\tilde{\zeta},\,\tilde{\pi}$ instead of $\zeta,\,\pi$. 
All the expressions in the following consideration are in the terms of $\tilde\zeta$ and $\tilde\pi$. However, the tilde above $\zeta$ and $\pi$ will be omitted for convenience, as there is no scope for confusion. So, whenever $\zeta,\,\pi$
appears below, the reader must keep in mind that they are actually the variables $\tilde{\zeta},\,\tilde{\pi}$ defined in Eq.~\eqref{simtransf}.

Using the above set of perturbation variables and performing several integrations by part,
we can write the action in the form
\be
S_2 = \int dt d^3 x \nb a^3 \cL_2\,,
\label{act3}
\ee
where
\ba
\cL_2 
&=&
A_1 \dot{\pi}^2 + A_2 \zeta \dot{\pi} + A_3 \zeta^2 + A_4 \pi^2 + A_5 \zeta \pi
\nonumber\\
&& 
+ A_6 \pi_{,i} \pi^{,i} + A_7 \zeta_{,i} \zeta^{,i} + A_8 \zeta_{,i} \pi^{,i}
\nonumber\\
&&
+ A_9 \dot{\zeta} \zeta_{, i}^{, i} + A_{10} \dot{\pi} \zeta_{, i}^{, i} 
+ A_{11} \dot{\zeta} \pi_{, i}^{, i} + A_{12} \dot{\pi} \pi_{, i}^{, i}
+ A_{13} \pi_{, i}^{, i} \zeta_{, i}^{, i}\,. 
\label{lag-2}
\ea
In the above, the coefficients $A_i\,$s depend on the background quantities. There expressions are large and it is not really necessary for our purpose to provide their explicit forms.
It follows from the above action that there is no kinetic term for $\zeta$,
which suggests that only $\pi$ is the propagating scalar degree of freedom in this theory. Since the gauge-invariant variable $\pi$ was originally defined in the same way as the scalar field perturbation in the uniform curvature gauge, clearly this propagating scalar degree of freedom comes from the scalar field sector, whereas gravity propagates only tensorial degrees of freedom. 

To eliminate the high-order spatial derivatives in the final action,
we impose the additional condition
\be
c_{3} = - c_{3}'\,.
\ee
Applying this condition to Eq.~(\ref{lag-2}),
we get
\ba
\cL_2 
&=&
A_1 \dot{\pi}^2 + A_2 \zeta \dot{\pi} + A_3 \zeta^2 + A_4 \pi^2 + A_5 \zeta \pi
\nonumber\\
&& + A_6 \pi_{,i} \pi^{,i} + A_7 \zeta_{,i} \zeta^{,i}+ A_8 \zeta_{,i} \pi^{,i}\,. 
\ea
The coefficients in the above equation can be written in terms of the dimensionless variables defined as
\ba
\epsilon &\equiv& - \frac{\dot{H}}{H^2}\,, \quad
\eta \equiv \frac{\ddot\phi}{\dot\phi H}\,,\quad
\epsilon_b \equiv \frac{\dot{b}_2}{b_2 H}\,,
\nonumber\\
\epsilon_x &\equiv&  \frac{X}{H^2}\,,\quad
\epsilon_{(1)} \equiv \frac{\dot\epsilon}{\epsilon H}\,,\quad
\eta_{(1)} \equiv \frac{\dot\eta}{\eta H}\,,\quad
\epsilon_{b (1)} \equiv \frac{\dot\epsilon_b}{\epsilon_b H}\,,
\ea
where some of them are slow-roll parameters in an inflationary scenario.
In terms of the above dimensionless variables,
we can write
\ba
A_1 &=& \frac{3 (b_2 - 3 \Omega_2) (1 + 2 \epsilon_x)^2}{4 (3 b_2 - 9 \Omega_2 -  \epsilon_x) \epsilon_x}\,,
\\
A_2 &=& - \frac{3 (b_2 - 3 \Omega_2) \epsilon H^3 (1 + 2 \epsilon_x)}{(-3 b_2 + 9 \Omega_2 + \epsilon_x) \dot{\phi}}\,,
\\
A_3 &=& \frac{3 \epsilon H^2 \bigl(b_2 (-3 + \epsilon) - 3 \Omega_2 (-3 + \epsilon) + \epsilon_x\bigr)}{6 b_2 - 2 (9 \Omega_2 + \epsilon_x)}\,,
\\
A_4 &=& \mbox{lengthy expression}\,,
\\
A_5 &=& - \frac{3 \epsilon H^3 \bigl(b_2 (3 + \epsilon + 2 \eta) - 3 \Omega_2 (3 + \epsilon + 2 \eta) -  \epsilon_x\bigr)}{2 (3 b_2 - 9 \Omega_2 -  \epsilon_x) \dot{\phi}}\,,
\\
A_6 &=& - \frac{1 - 8 c_{3} \epsilon_x}{4 a^2 \epsilon_x}\,,\quad
A_7 = \frac{2 c_{3} -  \epsilon_x}{2 a^2 \epsilon_x}\,,\quad
A_8 = - \frac{H + 4 c_{3} H}{a^2 \dot{\phi}}\,.
\label{coefs}
\ea
The terms $b_2'$ in all coefficients given by Eq.~(\ref{coefs}) are written in terms of the above dimensionless variables using Eqs.~(\ref{eom-b1}) and (\ref{eom-b2}) as
\be
3 b_2' = 2 \epsilon_x - 2 \epsilon_b b_2 + 2 b_2 \epsilon\,.
\label{b2p-slow}
\ee
Since $\zeta$ is a non-propagating degree of freedom,
it must be expressed in terms of $\pi$.
Varying the action in Eq.~(\ref{act3}) with respect to $\zeta$,
we obtain
\be
A_2 \dot{\pi} + 2 A_3 \zeta + A_5 \pi - 2 A_7 \zeta_{, i}^{, i} - A_8 \pi_{, i}^{, i} = 0\,.
\label{eqz}
\ee
The general solution for  Eq.~(\ref{eqz}) could be expressed in the form
\be
\zeta = {\cal C}(t) \exp\(\vec{f}\cdot\vec{x}\) + \frac{1}{2A_7} \int_{\vec{x_1}}^{\vec{x_2}} d^3 \tilde{x} G\(\vec{x}, \vec{\tilde{x}}\)\(A_2 \dot{\pi} + A_5 \pi - A_8 \pi_{, i}^{, i}\)\,, 
\label{zt-solution}
\ee
where ${\cal C}$ is a function of time, $\vec{f} = \left(\sqrt{\frac{A_3}{A_7}}, \sqrt{\frac{A_3}{A_7}}, \sqrt{\frac{A_3}{A_7}}\right)$ and $G$ is a green function.
The first term on the Right--Hand--Side  of Eq.~(\ref{zt-solution}) could be
interpreted as a gauge degree of freedom of $\zeta$,
while the second term indicates that $\zeta$ is non-local.
To compute the simplified relation between $\zeta$ and $\pi$
we assume that the background evolves like cosmic inflation, i.e., $\epsilon \ll 1$ and $\epsilon_{(1)} \ll 1$. 
We expand coefficients $A_2, A_3$  and $A_5$ up to the lowest order in slow-roll parameters $\epsilon$ as
\ba
A_2 &\simeq&  - \frac{3 (b_2 - 3 \Omega_2) \epsilon H^3 (1 + 2 \epsilon_x)}{(-3 b_2 + 9 \Omega_2 + \epsilon_x) \dot{\phi}}\,,
\\
A_3 &\simeq& - \frac{3}{2} \epsilon H^2\,,
\\
A_5 &\simeq& - \frac{3 \epsilon H^3 \bigl(b_2 (3 + 2 \eta) - 3 \Omega_2 (3 + 2 \eta) -  \epsilon_x\bigr)}{2 (3 b_2 - 9 \Omega_2 -  \epsilon_x) \dot{\phi}}\,. 
\ea
We see that  for the region with a physical  radius smaller than $1 / (\sqrt\epsilon H)$,
the spatial derivatives   terms in Eq.~(\ref{eqz}) are dominant,
so that we can write $\zeta$ in terms of $\pi$ as
\be
\zeta_{, i}^{, i} = r \pi_{, i}^{, i} 
= \frac{(H + 4 c_{3} H) \epsilon_x }{(2 c_{3} -  \epsilon_x) \dot{\phi}}\pi_{, i}^{, i}
\rightarrow
\zeta = r \pi + g(t)\,,
\ee
where $r=\frac{(H + 4 c_{3} H) \epsilon_x }{(2 c_{3} -  \epsilon_x) \dot{\phi}}$ and $g$ is a generic function of the time $t$ which will be set to zero.
Using the above relation and assuming that $\nb = a$, 
we can write the action in the form
\be
S = \int d\tau\{\(\frac{\partial v}{\partial \tau}\)^2 + M^2 v^2 - c_S^2 v_{, i} v^{, i} \}\,,
\label{s-inf}
\ee
where $\tau = \int da / a$ is the conformal time, $v \equiv \frac{\pi}{z}$
and 
\ba
z &\equiv & a \sqrt{A_1}\,,
\\
M^2 &\equiv & \frac{1}{z} \frac{d^2 z}{d \tau^2} + \frac{1}{z^2}\(- \frac{a^2}2 \overset{\bullet}{(r A_2)} + a^2 r^2 A_3 + a^2 A_4 + r A_5\)\,,
\label{m2}\\
c_S^2 &\equiv & - \frac{1}{z^2}\(A_6 + r^2 A_7 + r A_8\)
\nonumber\\
&=& - \frac{2 c_{3} (3 b_2 - 9 \Omega_2 -  \epsilon_x)}{3 (b_2 - 3 \Omega_2) (-2 c_{3} + \epsilon_x)}
= \frac{6 c_1 c_{3} + c_{3} \epsilon_x}{6 c_1 c_{3} - 3 c_1 \epsilon_x}\,,
\label{cs2-first} 
\ea
The expression for $M^2$ is lengthy,
so we will expand it in terms of the slow-roll parameters after we analyze a possible evolution of the background during inflation in the next section.

\section{Inflationary model}
\label{sec:6}

In this section, we will apply the results in the previous sections to study the dynamics of an inflationary universe.

\subsection{Background evolution}

To study the evolution of the background universe,
we integrate Eq.~(\ref{condition}) to get
\be
b_2' = - 2b_2 + B(t)\,,
\label{b2p-b2}
\ee
where $B(t)$ is a generic function of time.
Inserting Eq.~(\ref{b2p-b2}) into Eqs.~(\ref{eom-b1}) and (\ref{b2p-slow}),
we get
\ba
3 B H^2 - 9 b_2 H^2  &=& H^2 \epsilon_x + V\,,
\label{eomB1DL}\\
3 B - 6 b_2 &=& 2 b_2 \(\epsilon -  \epsilon_b\) + 2 \epsilon_x\,.
\label{eomB2Dl}
\ea
Eq.~(\ref{eom-b3}) can be written as
\be
\frac{d}{d t}\(X + V\) + 6 H X = 0\,,
\ee
so that Eq.~(\ref{eomB1DL}) yields
\be
2 B \epsilon - 6 b_2 \epsilon -  B \epsilon_B + 3 b_2 \epsilon_b - 2 \epsilon_x = 0\,,
\ee
where $\epsilon_B \equiv \dot{B} / (H B )$.
This equation gives
\be
\epsilon_b = - \frac{2 B \epsilon - 6 b_2 \epsilon -  B \epsilon_B - 2 \epsilon_x}{3 b_2}\,.
\label{epbSol}
\ee
Applying Eq.~(\ref{epbSol}) into Eq.~(\ref{eomB2Dl}),
we get
\ba
\epsilon_x &=& 3 b_2 (-3 + \epsilon) + B (\frac{9}{2} - 2 \epsilon + \epsilon_B)\,,
\label{ep-epx}\\
\epsilon_b &=& -6 + 4 \epsilon + \frac{B (3 - 2 \epsilon + \epsilon_B)}{b_2}\,.
\label{ep-epb}
\ea
Using Eqs.~(\ref{ep-epx}),
Eq.~(\ref{eomB2Dl}) yields
\be
V = - \frac{1}{2} \bigl(6 b_2 \epsilon + B (3 - 4 \epsilon + 2 \epsilon_B)\bigr) H^2\,. 
\label{ep-V}
\ee
We see that we can express $\epsilon_x, \epsilon_b$ and $V$ in terms of $\epsilon, \epsilon_B, B$ and $b_2$.
The variable $\eta$ can be expressed in terms of the dimensionless variables by writing Eq.~(\ref{eom-b3}) as
\ba
\eta &=& -\frac{\dot{V}}{2 H^3 \epsilon_x} - 3\,,
\nonumber\\ 
&=& \frac{6 b_2 \bigl(18 + (-12 + \epsilon_{(1)}) \epsilon + 2 \epsilon^2\bigr) -  B \bigl(54 + 4 \epsilon^2 + 9 \epsilon_B - 2 \epsilon_{B (1)} \epsilon_B - 2 \epsilon_B^2 + 2 \epsilon (-18 + 2 \epsilon_{(1)} + \epsilon_B)\bigr)}{2 \bigl(6 b_2 (-3 + \epsilon) + B (9 - 4 \epsilon + 2 \epsilon_B)\bigr)}\,,\label{ep-et}\nonumber\\
&&
\ea
where $\epsilon_{B (1)} \equiv \dot\epsilon_B / (H \epsilon_B)$.
Differentiating Eqs.~(\ref{ep-epb}) and (\ref{ep-et}) with respect to time,
we can write
\ba
\epsilon_{b (1)} &=& \frac{4 b_2^2 \epsilon_{(1)} \epsilon -  B^2 (3 - 2 \epsilon + \epsilon_B)^2 + B b_2 \Bigl(18 + 8 \epsilon^2 + (9 + \epsilon_{B (1)}) \epsilon_B + \epsilon_B^2 - 2 \epsilon \bigl(\epsilon_{(1)} + 3 (4 + \epsilon_B)\bigr)\Bigr)}{b_2 \bigl(2 b_2 (-3 + 2 \epsilon) + B (3 - 2 \epsilon + \epsilon_B)\bigr)}\,,
\label{ep-epbp}\\
\eta_{(1)} &=&  - \frac{6 b_2 \bigl(18 + (-18 + \epsilon_{(1)}) \epsilon + 4 \epsilon^2\bigr) -  B \bigl(54 + 4 \epsilon_{(1)} \epsilon + 12 \epsilon^2 + (9 - 2 \epsilon_{B (1)}) \epsilon_B - 2 \epsilon_B^2 - 2 \epsilon (27 + \epsilon_B)\bigr)}{6 b_2 (-3 + \epsilon) + B (9 - 4 \epsilon + 2 \epsilon_B)} \nonumber\\
&& + \frac{6 b_2 \Bigl(-108 + \bigl(144 + (-18 + \epsilon_{(2)}) \epsilon_{(1)} + \epsilon_{(1)}^2\bigr) \epsilon + (-60 + 8 \epsilon_{(1)}) \epsilon^2 + 8 \epsilon^3\Bigr)}{6 b_2 \bigl(18 + (-12 + \epsilon_{(1)}) \epsilon + 2 \epsilon^2\bigr) -  B \bigl(54 + 4 \epsilon^2 + 9 \epsilon_B - 2 \epsilon_{B (1)} \epsilon_B - 2 \epsilon_B^2 + 2 \epsilon (-18 + 2 \epsilon_{(1)} + \epsilon_B)\bigr)}\, \nonumber\\
&& + \frac{B \Bigl(324 - 24 \epsilon^3 + \bigl(54 + (-9 + 2 \epsilon_{B (2)}) \epsilon_{B (1)} + 2 \epsilon_{B (1)}^2\bigr) \epsilon_B + (-9 + 6 \epsilon_{B (1)}) \epsilon_B^2 + 2 \epsilon_B^3}{6 b_2 \bigl(18 + (-12 + \epsilon_{(1)}) \epsilon + 2 \epsilon^2\bigr) -  B \bigl(54 + 4 \epsilon^2 + 9 \epsilon_B - 2 \epsilon_{B (1)} \epsilon_B - 2 \epsilon_B^2 + 2 \epsilon (-18 + 2 \epsilon_{(1)} + \epsilon_B)\bigr)}\, \nonumber\\
&& + \frac{\epsilon^2 (180 - 20 \epsilon_{(1)} + 8 \epsilon_B) - 2 \epsilon \bigl(216 + (-27 + 2 \epsilon_{(2)}) \epsilon_{(1)} + 2 \epsilon_{(1)}^2 + (18 + \epsilon_{B (1)}) \epsilon_B + \epsilon_B^2\bigr)\Bigr)}{6 b_2 \bigl(18 + (-12 + \epsilon_{(1)}) \epsilon + 2 \epsilon^2\bigr) -  B \bigl(54 + 4 \epsilon^2 + 9 \epsilon_B - 2 \epsilon_{B (1)} \epsilon_B - 2 \epsilon_B^2 + 2 \epsilon (-18 + 2 \epsilon_{(1)} + \epsilon_B)\bigr)} \,. \nonumber\\
&& 
\label{ep-etp}
\ea
Let us analyze the dynamics of the background universe according to the above equations.
It follows from Eq.~(\ref{ep-epx}) that the evolution of the kinetic term of the scalar field depends on both $\epsilon$ and $\epsilon_B$.
In contrast to usual slow-roll inflationary models,
the condition $\epsilon \ll 1$ can be achieved even though $\epsilon$ is not small, i.e.,
$ 2 \epsilon_x \sim 9 B - 18 b_2 + 2 B \epsilon_B$ .
Since $B$ is a constant of integration,
there is no evolution equation for $B$.
However, it is possible to roughly determine how the evolution of $B$ influences the dynamics of the universe.
For a given value of $\epsilon_x $,
Eq.~(\ref{ep-epx}) gives
\be
\epsilon_B = - \frac{9 B - 18 b_2 - 4 B \epsilon + 6 b_2 \epsilon - 2 \epsilon_x}{2 B}\,.
\label{epB-s1}
\ee
Differentiating this equation with respect to time and dividing the result by $H \epsilon_B$,
we get
\ba
&& \epsilon_{B (1)} = \nonumber\\
&& \frac{B^2 \bigl(-27 + (9 + 4 \epsilon_{(1)}) \epsilon \bigr) - 2 \bigl(-3 b_2 (-3 + \epsilon) + \epsilon_x\bigr)^2 
+ B \bigl(3 b_2 (45 - 21 \epsilon - 2 \epsilon_{(1)} \epsilon + 2 \epsilon^2) + (27 + 2 \epsilon_{x (1)} - 10 \epsilon) \epsilon_x\bigr)}
{B \bigl(-6 b_2 (-3 + \epsilon) + B (-9 + 4 \epsilon) + 2 \epsilon_x\bigr)}\,,\nonumber\\
&&
\label{depB-s1}
\ea
where $\epsilon_{x (1)} \equiv \dot{\epsilon}_x / (H \epsilon_x)$.
   Substituting Eq.~(\ref{epB-s1}) and (\ref{depB-s1}) into Eq.~\eqref{ep-et},
   we get
\be
\eta =    \frac{1}{2} \(\epsilon_{x (1)} - 2 \epsilon\)\,.
\label{et-epx-ep}
\ee
The relation in Eq.~(\ref{et-epx-ep}) can directly be computed from the definitions of $\epsilon, \epsilon_x$ and $\eta$.
In usual inflationary models, $\epsilon_{x (1)} = \epsilon_{(1)}$.
Nevertheless, for this model,  $\epsilon_{x (1)}$ depends also on evolution of $B$,
i.e., the changing rate of the kinetic-energy fraction of the scalar field depends on both the changing rate of $\epsilon$ and $\epsilon_B$.
Hence, precise dynamics of inflation can be determined if the evolution of $B$ is known.
In the simple case,
we suppose that $B$ is nearly constant such that $\epsilon_B \ll \epsilon$ and $\epsilon_{B (1)} \ll \epsilon_{(1)}$.
At the lowest order in $\epsilon$,
Eq.~(\ref{ep-epb})  can be approximating as $\epsilon_b \simeq -6 + 3 B / b_2$ where $B$ is assumed to be constant in this approximation.
Hence, we get
\be
b_2 = C e^{-6 N} + \frac{B}2\,, 
\label{attr}
\ee
where $C$ is constant and $N \equiv \ln a$.
We see that $b_2 \to B /2$ exponentially,
so that $b_2 = B/2$ is the attractor of the background evolution.
At the attractor,
we have
\ba
\epsilon_x &=& - \frac{1}{2} B \epsilon \,,
\quad
\epsilon_b = 0\,,
\quad
\epsilon_{b (1)} = \frac{2 \bigl(9 + (-12 + \epsilon_{(1)}) \epsilon + 4 \epsilon^2\bigr)}{-3 + 2 \epsilon}\,.
\nonumber\\
\eta &=& \frac{1}{2} (\epsilon_{(1)} - 2 \epsilon)\,,
\quad
\eta_{(1)} = \frac{\epsilon_{(1)} (\epsilon_{(2)} - 2 \epsilon)}{\epsilon_{(1)} - 2 \epsilon}\,,
\quad
\frac{3}{2} B H^2 + H^2 \epsilon_x + V = 0\,.
\label{eps-att}
\ea
The relations in Eq.~(\ref{eps-att}) become the relations in the Einstein gravity if $B = - 2$.

Substituting Eq.~(\ref{ep-epx}) for $\epsilon_x$ into Eq.~(\ref{cs2-first}),
and expanding up to the lowest order in slow-roll parameters,
we get
\be
c_S^2 \simeq \frac{\bigl(3 B - 2 (b_2 + 6 c_{2})\bigr) c_{3}}{(b_2 - 3 c_{2}) (-9 B + 18 b_2 + 4 c_{3})}\,.
\label{cs2-slow}
\ee
According to $b_2 = B / 2$ at the attractor,
$b_2$ and $B$ have the same sign.
The range of parameters in which $c_S^2$ is positive can be $12 c_{2} < 3 B - 2 b_2$ and $b_2 > B/2$  for possitive $B$
while $12 c_{2} < 3 B - 2 b_2$ and $b_2 < B/2$ for negative $B$.
We always suppose that $c_{3}$ and $C_1 = b_2 - 3 c_{2}$ are possitive.
At the attractor,
we have $c_S^2 = 1$.

Using Eqs.~(\ref{ep-epx}), (\ref{ep-epb}), (\ref{ep-et}) - (\ref{ep-etp}),
we can write Eq.~(\ref{m2}) up to the lowest order of $\epsilon$ as
\ba
M^2 &\simeq & - \frac{81 B^4 - 108 B^3 (5 b_2 + 3 c_{2}) - 72 B^2 (28 b_2^2 - 303 b_2 c_{2} + 414 c_{2}^2)}{16 (b_2 - 3 c_{2})^2 \bigl(3 B - 2 (b_2 + 6 c_{2})\bigr)^2 \tau^2}\,\nonumber\\
&+& \frac{48 B (341 b_2^3 - 2565 b_2^2 c_{2} + 4968 b_2 c_{2}^2 - 2484 c_{2}^3)}{16 (b_2 - 3 c_{2})^2 \bigl(3 B - 2 (b_2 + 6 c_{2})\bigr)^2 \tau^2}\,\nonumber\\
&-& \frac{16 (1385 b_2^4 - 10482 b_2^3 c_{2} + 24084 b_2^2 c_{2}^2 - 18360 b_2 c_{2}^3 + 2592 c_{2}^4)}{16 (b_2 - 3 c_{2})^2 \bigl(3 B - 2 (b_2 + 6 c_{2})\bigr)^2 \tau^2} \,,
\label{m2-slow}
\ea
where we have used $aH = - 1 / (\tau (1 - \epsilon))$ for the quasi de Sitter expansion \cite{Riotto}.
When the background evolution reaches the attractor,
Eq.~(\ref{m2}) becomes
\be
M^2 \simeq \frac{2}{\tau^2} + \frac{3}{2 \tau^2} \(\epsilon_{(1)} - 2 \epsilon\)
= \frac{2}{\tau^2} + \frac{3}{\tau^2} \eta\,. 
\label{m2-att}
\ee

\subsection{Scalar perturbations}

To compute primordial perturbation during inflation,
we write $v$ in the form of the Fourier expansion as
\be
v({\bf x}, t) = \int\,\frac{d^3{\bf k}}{(2\pi)^{3/2}} e^{i{\bf k}\cdot{\bf x}} v_{{\bf k}}(t)\,,
\ee 
so that the evolution equation from the action (\ref{s-inf}) is
\be
\frac{d^2v_k}{d \tau^2} + \(c_S^2 k^2 - M^2\)v_k = 0\,.
\label{v-eom}
\ee
It follows from Eq.~(\ref{attr}) that the attractor for the background evolution is reached in a few numbers of e-folding for $C \sim 1$ and $B \sim 1$.
Hence, we concentrate on  the case where the background evolution already reaches the attractor,
so that $c_S^2$ is set to unity and Eq.~(\ref{m2-att}) is used.
We can write Eq.~(\ref{v-eom}) as
\be
\frac{d^2v_k}{d \tau^2} + \(k^2 - \frac{1}{\tau^2}\(\mu^2 - \frac 14\)\)v_k = 0\,,
\label{eom-v-bes}
\ee
where
\be
\mu^2 \equiv \frac{3}{4} (3 + 4 \eta )\,. 
\ee
The solutions for eq.~(\ref{eom-v-bes}) are in the form of Bessel functions.
However, we consider the solution in the cases where the wavelength of the perturbation modes is much smaller or much larger than the Hubble radius $H^{-1}$.
For the short wavelength  modes,
the minimal quantum fluctuations state provides the normalized positive frequency modes as \cite{Garriga}
\be
v_{k} \approx {\frac{e^{-ik \tau }}{(2 k)^{1/2}}}\,,
\quad \mbox{for} \quad
(aH\ll  k)\,.
\label{short-sol}
\ee
For the long wavelength mode,
Eq.~(\ref{eom-v-bes}) becomes
\be
\frac{d^2v_k}{d \tau^2} - \frac{1}{\tau^2}\(\mu^2 - \frac 14\)v_k = 0\,.
\label{eom-v-long}
\ee
The solutions for the above equation are
\be
v_k = c(k) \tau^{1/2 \pm \mu}\,,
\quad \mbox{for} \quad
(aH\gg k)\,.
\label{long-sol}
\ee
where $c(k)$ is a function of the wavenumber $k$.
In the  following analysis,
we will consider only the dominant solution $v_k \propto \tau^{1/2 + \mu}$ only.
Matching Eq.~(\ref{short-sol}) with Eq.~(\ref{long-sol}) at the time of horizon crossing ($aH=k$).,
we get $|c(k)|^2 =  k^{2\mu} / 2$ and therefore we obtain the solution at the moment $aH=k$ as
\be
|v_k|^2 = \left. \frac 12 k^{2 \mu} \tau^{1 + 2 \mu}\right|_{aH = k}
= \left. \frac{1}{2 a H}\right|_{aH = k}\,.
\label{v2-sol}
\ee
To connect with the observable quantities,
we write the above solution in terms of the curvature perturbation defined in a similar way as in GR as 
\be
\xi_k = - \frac{H}{\dot{\phi}} \pi_k
= - \frac{H}{\dot{\phi}} \frac{v_k}{z}
= - \frac{1}{\sqrt{2 \epsilon_x}} \frac{v_k}{z}\,.
\ee
Hence, Eq.~(\ref{v2-sol}) yields
\be
|\xi_k|^2
= \left. \frac{1}{4 \epsilon_x a H z^2}\right|_{aH = k}\,.
\ee
The power spectrum for the scalar perturbation is
\be
{\cal P}^{(S)} \equiv \frac{1}{2\pi^2} k^3 |\xi_k|^2
= \frac{1}{8\pi^2} \left. \frac{H^2}{\epsilon_x A_1}\right|_{aH = k}
\label{powerS}
\ee
The corresponding spectral index is
\ba
n_S - 1 &\equiv& \frac{d\ln {\cal P}^{(S)}}{d\ln k}
= \frac 1{H} \frac{d\ln {\cal P}^{(S)}}{d t}
\nonumber\\
&=& -2 \epsilon - \frac{\dot\epsilon_x}{H \epsilon_x} - \frac{\dot{A}_1}{H A_1}
= -2 \epsilon + {\cal O}(\epsilon^2)\,.
\label{ns-1}
\ea
We note that the term $\dot\epsilon_x / (H \epsilon_x) + \dot{A}_1 / (H A_1) \sim {\cal O}(\epsilon^2)$.

\subsection{Tensor perturbations}

For the tensor modes,
the metric perturbation can be decomposed as
\be 
h_{ij} = a^2\(\delta_{ij} + \gamma_{ij}\)\,,
\qquad
h^{ij} = a^{-2}\(\delta^{ij} - \gamma^{ij}\)\,,
\label{tensor:met}
\ee
where $\gamma_i^i = 0$ and $\partial_i\gamma^{ij} = 0$.
Since the gauge-fixing condition in Eq.~(\ref{lg}) depends on the trace of $K_{ij}$,
it has no effect on the tensor perturbations.
From the Lagrangian in Eq.~(\ref{lang}),
we can write the second order action for the tensor perturbations  as
\be
S_T = \int dt dx^3 a^3 \bar{N}\(
\frac{c_1}{4} \dot{\gamma}_{ij} \dot{\gamma}^{ij}
- \frac{c_{3}}{4 a^2} \partial_i \gamma^{kl} \partial^i \gamma_{kl}
\)\,.
\label{act:tensor1}
\ee
The tensor perturbation $\gamma_{ij}$ can be expanded in terms of the polarization tensors as \cite{Maldacena, Guzzetti}
\be
\gamma_{ij} = \int {d^3k \over ( 2 \pi)^3} \sum_{s=\pm} \epsilon_{ij}^s(k)
\gamma^s_{k}(\tau) e^{ i \vec k \cdot \vec x }\,,
\ee
where $\epsilon_{i i} = k^i \epsilon_{ij} =0$ and $\epsilon^s_{ij}(k)
\epsilon^{s'}_{ij}(k)  = 2 \delta_{s s'}$.
We define \cite{Guzzetti}
\be
v^s_T \equiv z_T \gamma^s_k\,,
\quad\mbox{where}\quad
z_T \equiv a \sqrt{\frac{c_1}{2}}\,.
\ee
Setting $\bar{N} = a$, the action (\ref{act:tensor1}) yields the following equation of motion for $v^s_T$
\be
\frac{d^2 v^s_T}{d \tau^2} + \(c_T^2 k^2 - \frac{d^2 z_T}{z_T d \tau^2}\)v^s_T = 0\,,
\label{eom-vt}
\ee
where the sound speed square of the tensor mode is $c_T^2 \equiv c_{3} / c_1$.
Deep inside the sound horizon,
the normalized positive frequency modes of $v^s_T$ are
\be
v^s_T \approx {\frac{e^{-ik c_T\tau }}{(2c_T k)^{1/2}}}\,.
\ee
On the super horizon scales,
we have $v^s_T = T[k) z_T$,
where $T(k)$ is a function of $k$.
Matching the long wavelength with the short wavelength solutions,
the expression for $v^s_T$ at the moment  of sound horizon crossing is
\be
|\gamma^s|^2 = \left. \frac{1}{2 c_T k z_T^2} \right|_{a H = k c_T}\,. 
\label{vt-super}
\ee
The power spectrum for tensor perturbations is defined as
\be
{\cal P}^{(T)} 
\equiv \frac{k^{3}}{2\pi^{2}}\sum_{s=\pm}\(\left| \gamma^s\right|^2 \)\,.
\label{pktensor}
\ee
Hence, at the moment  of sound horizon crossing
Eq.~(\ref{vt-super}) yields 
\be
{\cal P}^{(T)} =
\left. \frac{H^2}{\pi^2 c_T^3 c_1} \right|_{a H = k c_T}\,.
\ee
The tensor--to--scalar ratio is given by
\be
r \equiv \frac{{\cal P}^{(T)}}{{\cal P}^{(S)}}
= \frac{8 \epsilon_x  A_1}{c_T^3 C1}
\sim 
\frac{2 c_1^2}{c_{3}^3} 
+ \frac{B c_1^2 (1 - 12 B + 72 c_{2}) \epsilon}{3 (B - 6 c_{2}) c_{3}^3}\,.
\label{r}
\ee
Imposing the condition $c_T^2  = 1$ according to the constraint on propagation speed of gravitational waves \cite{Monitor:2017mdv},
we get
\be
r \sim \frac{2 }{c_{3}}  + {\cal O}(\epsilon)\,,
\ee
which indicates that $r \propto 1/ c_{3}$.
In contrast to the usual slow-roll inflationary model in which $r = 16\epsilon$,
the ratio $r$ can either be small or be in the order of unity depending on the value of $c_{3}$.

\section{Conclusions}\label{sec:7}

In this work, we investigate a the Spatially Covariant Gravity using perturbative analysis.
We extend the analysis in \cite{Hu:2021yaq} by adding a scalar field in the theory, to make it more relevant in cosmology. 
Using the perturbative analysis technique similar to \cite{Hu:2021yaq}, we deduce the condition for the gravitational sector to propagate only two tensorial degrees of freedom and no additional scalar degree of freedom, up to the lowest perturbative order. The condition obtained is different from the one for vacuum, obtained in \cite{Hu:2021yaq} We then use the resulting theory with a scalar field to describe the inflationary universe and compute the observational predictions from the model. 

Similar to the usual MMG theories, a gauge-fixing is required to keep having two tensorial degrees of freedom for gravity when the matter field appears in the theories. Inspired by the gauge-fixing condition used in \cite{Aoki:2018zcv,Aoki:2020oqc} the Hamiltonian formalism of MMG theories, i.e. $f(\mathcal{H})$ theories, propose the gauge-fixing condition in the Lagrangian formalism, and investigate the condition for which the spatially covariant theory of gravity in our consideration has two tensorial degrees of freedom up to the first order in cosmological perturbations. 
We make this explicit by writing the second order action for perturbations solely in terms of one single gauge-invariant scalar perturbation quantity that is defined in the same way as the scalar field perturbation in the uniform 3-curvature gauge (Eq.~\eqref{s-inf}).

We apply the resulting actions for the first and second orders perturbations to cosmic inflation.  
Analyzing the background dynamics, we are able to find an attractor solution. The attractor background is then used to investigate scalar and tensor perturbations. At the attractor solution, we find that the sound speed squared for scalar perturbation is equal to unity. The scalar spectral index solely depends on the Hubble slow-roll parameter $\epsilon\equiv-\frac{\dot{H}}{H^2}$. Demanding that the primordial gravitational waves propagate at the speed of light, the tensor-to-scalar ratio is inversely proportional to the coefficient $c_3$, and can therefore be in the order of unity or small depending on $c_3$.

One may argue that the condition \eqref{condition} we obtained using the perturbative technique does not really that the resulting theory an MMG, as the scalar mode that we killed at the lowest order in perturbation can again show up in higher orders. Exact conditions for the absence of the extra gravitational scalar degree of freedom can only be calculated by a Hamiltonian analysis, where we obtain a degeneracy condition and a consistency condition \cite{Gao:2019twq}. Indeed, this point has been recognized in earlier works on perturbative construction of MMG theories starting from SCG theories \cite{Gao:2019lpz,Hu:2021yaq}. If one takes into account expansion the action up to the cubic order in perturbation around an FLRW background, one will get additional conditions on the coefficients \cite{Hu:2021yaq}. Ideally, one needs to perform the same analysis order by order and ensure killing of the unwanted degree of freedom at each order. In this sense, the condition \eqref{condition} that we obtain is only an approximate condition that is necessary, but not sufficient for the theory to be an MMG. However, as long as we confine ourselves in the domain of linear perturbations, which is what we mostly work with in cosmology, this approximate condition is good enough for calculating observable quantities. 

\section*{Acknowledgments}

This research has  received funding support from the NSRF via the Program Management Unit for Human Resources and Institutional Development, Research and Innovation [grant number B01F650006].

\end{document}